\documentclass[twocolumn,preprint,5p]{elsarticle}
\usepackage{epsfig}
\usepackage{amssymb}
\usepackage{lineno}
\journal{Astroparticle Physics}
\def\hess{H.E.S.S.}

\def\gr{$\gamma$-ray}
\def\grs{$\gamma$~rays}

\def\x{$\times$}

\def\mp{$\pm$}

\def\deg{$^{\circ}$}

\def\diff{cm$^{-2}$s$^{-1}$TeV$^{-1}$}
\def\integ{cm$^{-2}$s$^{-1}$}

\begin{document}

\begin{frontmatter}

\title{Optimization of multivariate analysis for IACT stereoscopic systems}

\author[lapp]{A.~Fiasson\corref{cor}}
\ead{fiasson@in2p3.fr}
\cortext[cor]{Corresponding author.}
\author[lapp]{F.~Dubois}
\author[lapp]{G.~Lamanna}
\author[lapp]{J.~Masbou}
\author[lapp]{S.~Rosier-Lees}
\address[lapp]{LAPP, Laboratoire d'Annecy-le-Vieux de Physique des Particules, UMR/IN2P3-CNRS, Université de Savoie, 9 Chemin de Bellevue - 74941 Annecy-le-Vieux, France}

\begin{abstract}

Multivariate methods have been recently introduced and successfully applied for the discrimination of signal from background in the selection of genuine very-high energy gamma-ray events with the H.E.S.S. Imaging Atmospheric Cerenkov Telescope. The complementary performance of three independent reconstruction methods developed for the H.E.S.S. data analysis, namely Hillas, model and 3D-model suggests the optimization of their combination through the application of a resulting efficient multivariate estimator. In this work the boosted decision tree method is proposed leading to a significant increase in the signal over background ratio compared to the standard approaches. The improved sensitivity is also demonstrated through a comparative analysis of a set of benchmark astrophysical sources.

\end{abstract}


\end{frontmatter}

\linenumbers

\section{Introduction}
In the past decade, a new astronomical window has been opened thanks to the last generation of ground-based Imaging Atmospheric Cerenkov Telescopes (IACTs). Before the construction of IACT arrays such as H.E.S.S., M.A.G.I.C. and V.E.R.I.T.A.S., only a few very-high energy (VHE) \gr\ sources ($>$100~GeV) were known. This new generation of experiments has resulted in the discovery of many tens of galactic and extra-galactic \gr\ sources.

The H.E.S.S. system is currently the most efficient instrument to look at the inner part of the Galactic plane. The system is composed of four IACTs and provides a sensitivity to a 1\% of Crab Nebula flux in around 25 h of observations~\cite{Vincent08}. A systematic survey of about a third of the Galactic plane has been conducted since the beginning of the observations in full operation mode in 2004, leading to  the discovery of more than 50 sources within our Galaxy~\cite{Aharonian05}\cite{Aharonian06c}.

IACTs detect the Cerenkov light emitted by the secondary particle showers generated by the interaction of the incoming \gr\ into the atmosphere. They face a dominant background due to the hadron induced showers in the research of \gr\ signal. Three alternative reconstruction and discrimination methods have been developed and applied so far to the H.E.S.S. data analysis, namely Hillas, model and 3D-model. They have been individually improved and updated in the last years. The Xeff multivariate analysis method has been recently introduced~\cite{Dubois09} in the \hess\ data analysis, increasing the discrimination power of genuine VHE gamma-ray event signals from the cosmic-ray background and improving the reconstruction performance (e.g. energy and direction reconstructions) through the combination of the three methods together. In this work the optimization of the multivariate analysis is presented through the application of a boosted decision tree (BDT) method leading to a single alternative discriminating estimator. After describing the methodology, some examples of application of the proposed multivariate method are presented in order to demonstrate the achieved gain in terms of sensitivity and precision.

\section{Current methods used in \hess\ data analysis}

The three shower reconstruction methods applied so far in the HESS data analysis are briefly described in this section.

\subsection{Hillas analysis}

This historical method has been introduced by M. Hillas in 1985 for single telescope analysis~\cite{Hillas85} and was the first method applied to the \hess\ data analysis for multi-telescope images~\cite{Aharonian06a} (hereafter Hillas method). The so called Hillas parameters of the shower are extracted by fitting an ellipse to the images. The dimensions (length and width) and orientation of the image on the focal plane (azimuthal angle, distance of the image barycenter to the camera center ...) are estimated from the fit. A charge measurement (in number of photo-electrons) is obtained from the total amplitude of the images. The direction of the incoming particle is estimated through the ellipse orientation, while the shower energy is estimated with the total image amplitude and the reconstructed impact parameter of the shower. The discrimination between hadron and \gr\ events is done through scaled variables. The Hillas geometric parameters of the image, the length and width, are scaled with the mean values and the dispersion obtained from Monte Carlo simulations. The scaled variables are then averaged over the various triggered telescopes. They are often combined in an unique discriminating variable mean scaled sum defined as $(\mathrm{Mean Scaled Width + Mean Scaled Length})/\sqrt{2}$. For more information see~\cite{Aharonian06a}.

\subsection{Semi-analytical model analysis}

This method was first developed by the CAT collaboration~\cite{LeBohec98} and has been applied to \hess\ data analysis~\cite{DeNaurois09} (hereafter model method). It is based on the comparison of the shower image with a shower prediction given by a semi-analytical model. The image is compared to the images stored in a model look-up table and a log-likelihood minimization of the fit is done over all the available pixels. The parameters of the most probable image give the primary particle energy and incoming direction. The discrimination between \grs\ and hadrons is achieved with a goodness-of-fit variable combined with the shower primary depth, which is a free parameter of the model. As for the Hillas analysis, this variable is often rescaled with the simulation mean value and dispersion. Recent improvements of this analysis method have greatly increased the sensitivity. For more information see~\cite{DeNaurois09}.

\subsection{3D model analysis}

The 3D-model reconstruction is the third method developed for the analysis of \hess\ data~\cite{Lemoine06} (hereafter model3D method). It consists of modeling the atmospheric shower as a Gaussian photosphere with anisotropic angular light distribution. This model is then used to predict the light collected in each pixel of the camera. Several shower parameters are extracted from the fit of the recorded image with the model prediction. The rotational symmetry of the shower with respect to the main axis can then be used to discriminate hadrons and \grs, through the reduced 3D-width variable. For more information see~\cite{Naumann09}.

\section{The Boosted Decision Tree method}

The discrimination methods previously described are all based on simple cut based analysis techniques. Extension of such techniques such as neural networks or decision trees, already used in a wide range of domains, have been introduced and applied in the field of high energy physics. They have the main advantage to consider non-linear correlations between input parameters. The decision trees have the particularity to be insensible to the use of parameters without discrimination power.

A decision tree is a decision support tool that uses a tree-like model of decision to separate two populations in terms of signal or background~\cite{Breiman84}. Starting from the initial event sample, a search for the best criterion among the discriminant variables is performed. The selection results in two event samples that are submitted to the same procedure. Repeated binary selections are then performed on the subsequent event samples until some stop criterion is reached. 
When the splitting is stopped, the events from the extremal folders (which are called leaves) are classified in terms of signal or background likeliness according to the class the majority of events belongs to. The stop criterion is set in order to avoid a too efficient discrimination between signal and background. The splitting could continue until the leaves contain only signal or background events, that would imply that the trees are overtrained. In order to avoid this problem, a pruning of the tree is necessary to remove the statistically insignificant nodes.

The boosting process aims to stabilize the response of a tree and improve its performance. The BDT method consists of a forest of successive trees. Misclassified events in the previous tree are given a higher event weight on the following tree. In the most popular boosting technique, AdaBoost~\cite{Freund95}, the following tree is trained with a modified initial event sample where the weight of misclassified events is multiplied by a boost weight $\alpha$. This weight is derived from the fraction of misclassified event $err$ on the previous tree as $\alpha = \frac{1-err}{err}$. Once the forest has been defined, the signal or background likelihood of individual events is estimated applying the set of splitting of the various trees, and is averaged over the forest according to weights, set to stabilize the decision procedure.

\begin{figure*}[!t]
  \centering
  \includegraphics[width=1.\textwidth]{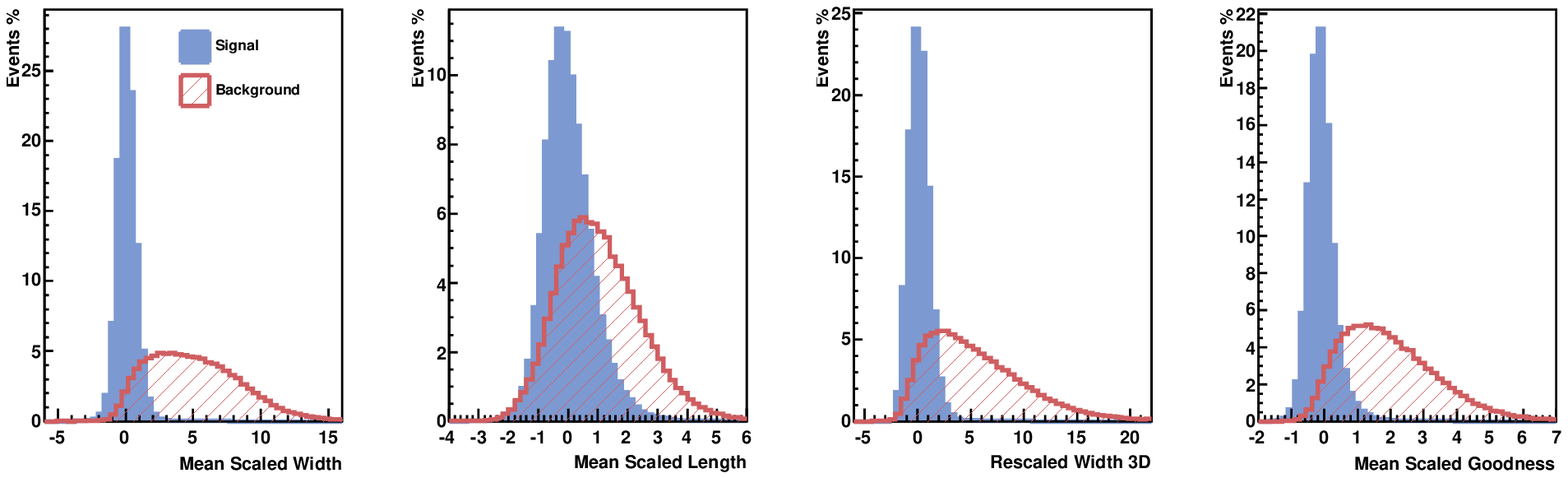} \\
  \includegraphics[width=1.\textwidth]{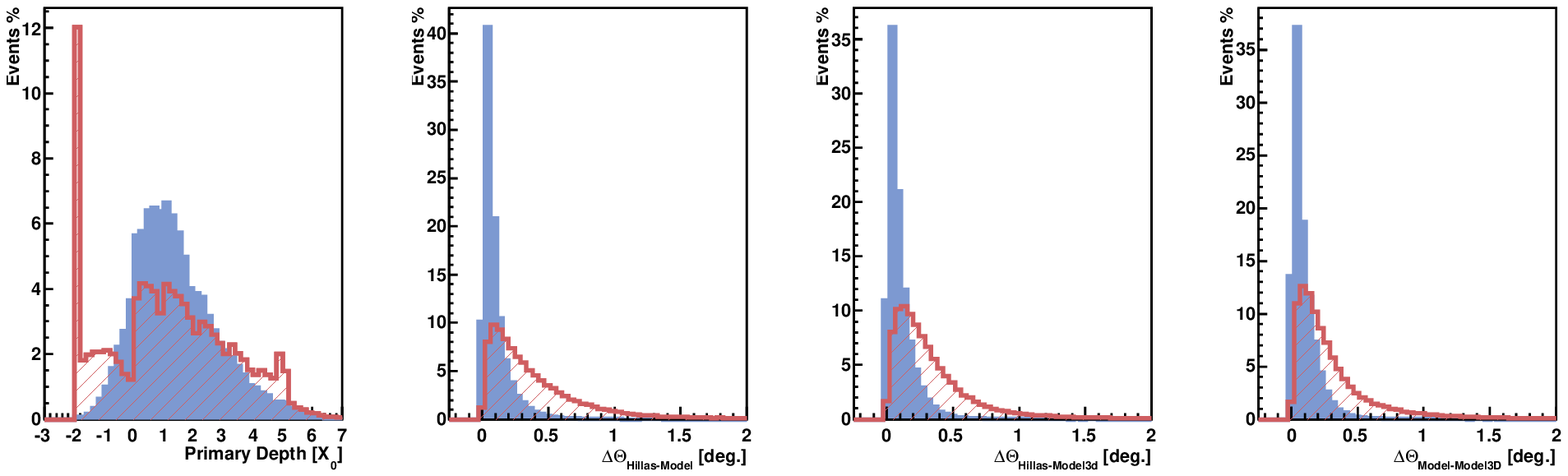}
  \caption{Distributions of the variables used in the BDT method for a zenith angle ranging between 25\deg\ and 35\deg\ and an energy between 500~GeV and 1~TeV. The blue filled and hatched red histograms are the simulated \gr\ event and background event distributions respectively. The four upper panels are the main variables coming from the original analysis methods: Mean Scaled Width, Mean Scaled Length, Rescaled Width 3D and Mean Scaled Goodness (from left to right). The lower panels include the additional variables: Primary Depth and the difference between reconstructed directions: Hillas, model \& model3D (left to right).}
  \label{fig1}
\end{figure*}

\section{Application to \hess\ data}

Boosted decision trees have already been applied for the analysis of \hess\ data to discriminate between showers generated by leptons and hadrons. It led to a ground-based measurement of the electron + positron spectrum with \hess~\cite{Aharonian08}. Another analysis has been developed with a combination of variables derived from the Hillas-moment method. It showed a clear improvement of the hadron-\grs\ discrimination compared to the standard Hillas analysis~\cite{Ohm09}.

The aim of this work is to apply the BDT technique to the various methods of Cerenkov shower image analysis currently used by the \hess\ Collaboration as described in section~2. The combination of these independent and complementary methods is expected to improve significantly the \gr\ hadron discrimination. In this section, the procedure followed in this work is described.

\subsection{Training samples}

The \gr\ event sample used to train the BDTs has been taken from Monte Carlo simulations. The \grs\ have been simulated through the shower simulation code KASKADE~\cite{Kertzman94}, with an impact parameter up to 550~meters from the array center. The zenith angle varies from 0\deg\ to 70\deg . The off-axis angle of the showers have been taken from 0\deg\ to  2.5\deg\ from the camera axis by steps of 0.5\deg\ . It corresponds to the actual field of view of the \hess\ camera. 

The background event sample has been selected from real \hess\ events. The events have been chosen from extra-galactic observations in order to avoid a contamination by a potential diffuse \gr\ background or undetected galactic \gr\ sources. The events coming from known extra-galactic VHE \gr\ sources have been excluded and the remaining events detected by \hess\ in those observations are considered as background events. Most of these showers are generated by hadron cosmic rays or electrons. The extra-galactic diffuse \gr\ background is usually assumed to be negligible at TeV energies.

\subsection{Input variables for combination}

Four variables are used: the mean scaled width and length of the images from the Hillas method; the mean scaled goodness from the model analysis; and the rescaled width from the model3D analysis. The distribution of the variables for simulated \gr\ and hadron events is shown on the upper panel of figure~\ref{fig1}. As expected, the \gr\ distributions are centered on the origin while the hadron distributions are shifted towards larger values. Moreover, it has been shown that these variables are almost not correlated for \grs\ and are partially correlated for hadrons~\cite{Dubois09}.

Additionally to these variables, a set of 4 variables has been added to improve the discrimination. The primary interaction depth of the particle, scaled in term of photon radiation length, has been shown to have a significant different distribution for hadrons and \grs ~\cite{DeNaurois09}. Figure~\ref{fig1} shows the distribution of the primary interaction depth for hadrons and \grs\ estimated through the model method. It should be noted that this variable is obtained by comparison to simulated \gr\ showers and assumes a \gr\ nature of the event. The non convergence of the fit procedure for the background is responsible for the overflow cumulative bin observed in figure~\ref{fig1} (at -2). This variable is already used to operate an event preselection by the model and model3D analyses. Moreover, each reconstruction method gives a reconstructed direction, which are not necessarily identical. It has been shown that the fluctuations of these reconstructions are bigger for hadrons than for \grs~\cite{Dubois09}. An additional discrimination can be thus achieved using the reconstruction differences looking at two methods alternatively ($\Delta\theta_{\mathrm{Hillas-Model}}$, $\Delta\theta_{\mathrm{Hillas-Model3D}}$ and $\Delta\theta_{\mathrm{Model-Model3D}}$). The distribution of the additional variables is shown in the lower panel of figure~\ref{fig1} for hadrons and \grs . The discrimination power of these variables is visible in this figure.

\subsection{Training strategy}

The shapes of the particle shower and the Cerenkov image change with the particle energy. The distribution of the discriminating variables changes as well. The four variables from the standard methods are scaled using look-up tables generated with MC simulations, as a function of energy and observing zenith angle. Although, since the variables are scaled, large variations with the energy and the observation conditions of their distribution are not expected. On the contrary, the second set of variables, including the primary depth and the difference in reconstructed directions, is dependent on the energy of the particle and the observation conditions.

The strategy followed to discriminate between hadrons and \grs\ is the same as in~\cite{Ohm09}. The \hess\ dynamical range has been divided in six energy bands from 100~GeV to 100~TeV. Both simulated and real events have been distributed within these bands using the energy derived by the combined method already applied in the XEff analysis and described in~\cite{Dubois09}. This method combines the reconstructed energy derived from the three original methods and improves the energy resolution compared to the single analysis. As well, the zenith angle range (0\deg -70\deg) has been divided into seven bands. The BDTs have been trained separately within these energy and zenith angle bands. Due to trigger effects, the statistics within several low energy bins at high zenith angle is very low and the corresponding events have been neglected: zenith angle larger than 35\deg\ and 52.5\deg\ respectively for the first and second energy bands (empty bins in figures~\ref{fig4} and~\ref{fig6}). The training and events selection have been done using the BDT method provided by the package for multivariate analysis TMVA~\cite{Hoecker07}. The adaptative boosting method AdaBoost has been used. 

Several parameters can be modified in order to improve the method efficiency and to control its stability. A particular attention has been brought to the control of the over-training of the BDT. A too efficient classifying tree can lead to bias effects. The BDT response has been controlled with an independent event sample. The consistency of the training and test sample BDT distributions has been checked for each zenith angle and energy bin. The parameters of the BDT trainings have been slightly modified with respect to the default values optimised by the TMVA developers. Their choice is the result of a compromise between hadron-\gr\ discrimination efficiency and the absence of over-training. The tree forests are composed of 200 trees. The selection splitting has been done at the node level performing 100 steps over the variable distributions. The separation between the populations is performed using the Gini Index criterion, defined as $p\times(1-p)$ where $p=\frac{S}{S+B}$ is the purity of the sample ($S$ and $B$ are the signal and background events). Further splitting has been stopped when the number of events fell below 20. The tree pruning is performed using the cost complexity method~\cite{Breiman84} with a pruning strength set at 20.

\subsection{BDT response}

\begin{figure}[!t]
  \centering
  \includegraphics[width=0.45\textwidth]{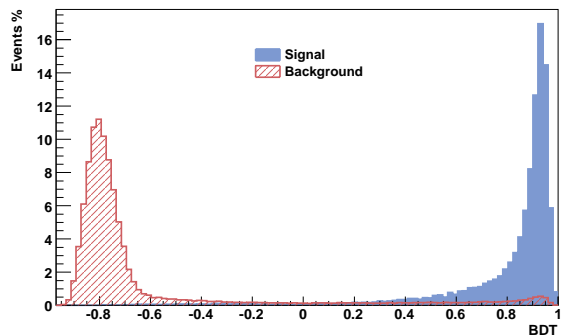}
  \caption{BDT response distribution for the events with a zenith angle ranging between 25\deg\ and 35\deg\ and a reconstructed energy between 500~GeV and 1~TeV. The blue filled and hatched red histograms are the simulated \gr\ and background distributions respectively.}
  \label{fig2}
\end{figure}

\begin{figure}[!t]
  \centering
  \includegraphics[width=0.45\textwidth]{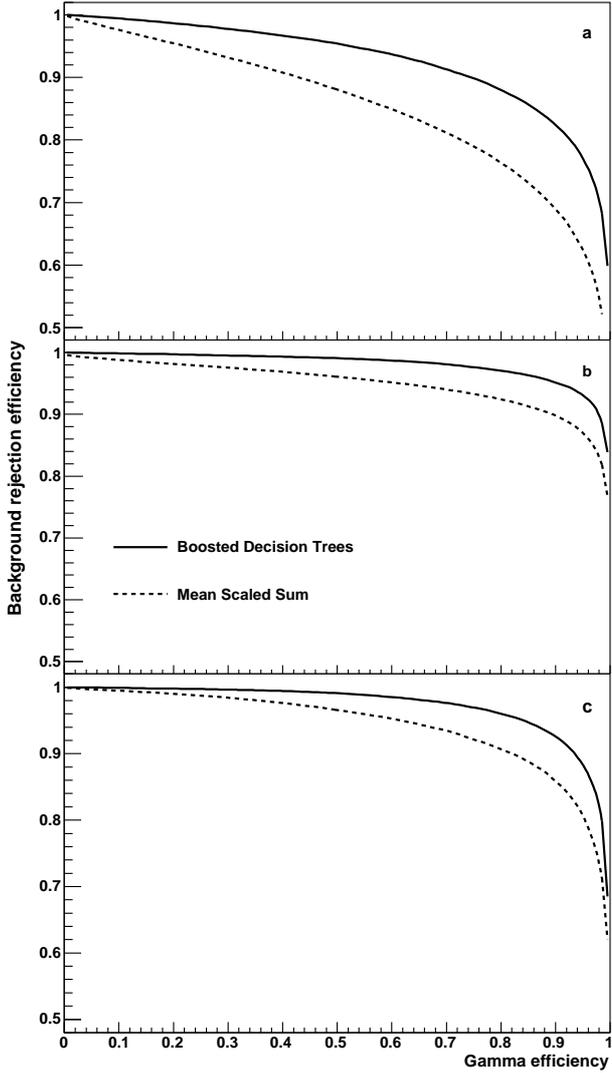}
  \caption{Background rejection efficiency as a function of the \gr\ efficiency (Receiver operator characteristic diagram). The BDT method presented in this paper is shown with the continuous line, while the mean scaled sum variable from the Hillas analysis is shown with the dotted line. The diagrams are for events with an energy and a zenith angle corresponding to: a) 100~GeV-300~GeV and 0\deg -15\deg\ b) 500~GeV-1~TeV and 25\deg -35\deg\ c) 500~GeV-1~TeV and 52.5\deg -70\deg .}
  \label{fig3}
\end{figure}

Figure~\ref{fig2} shows the results of the tests for the trained BDT with an independent test sample of \gr\ and background events for one zenith angle and energy band. The discrimination power of this new variable is clearly visible when compared to the distributions of the original variables. The rejection efficiency is clearly improved with respect to the original variables for all energies and zenith angle. Figure~\ref{fig3} shows for three case the receiver operator characteristic (ROC) diagram for the BDT method described in this paper compared to the main variable from the original Hillas method. For a given level of hadron rejection, the combined estimator allows to keep a more important fraction of \gr\ events. It shows the improvement in terms of hadron rejection possible through this combination of methods.

\subsection{Selection cuts determination}

Once the BDTs have been trained, a choice on the cut values on the estimator has to be made. Three optimization strategies have been applied depending on the source strength. The first set is dedicated to the analysis of strong sources such as the Crab Nebula. The second is defined for intermediate source fluxes of the order of 10\% the Crab Nebula flux. The last one is optimized for faint source searches with flux of the order of 1\% the Crab Nebula flux. The cut values have been chosen for these three sets within each zenith angle and energy band. The Crab Nebula has been used as a reference. The signal over background ratio has been estimated without any event selection within a region of 0.11\deg\ around the position of the Nebula (standard angular cut for point-like source). For every value of the BDT estimator, the corresponding \gr\ and background event efficiencies have been applied to 100\%, 10\% and 1\% of the measured Crab Nebula signal over background ratio, for each set of cuts respectively. In each band, the BDT output value for which the significance of the signal $S/\sqrt{S+B}$ is maximum has been chosen. $S$ and $B$ are in this formula the events selected from the signal and background samples.

\begin{figure}[!t]
  \centering
  \includegraphics[width=0.45\textwidth]{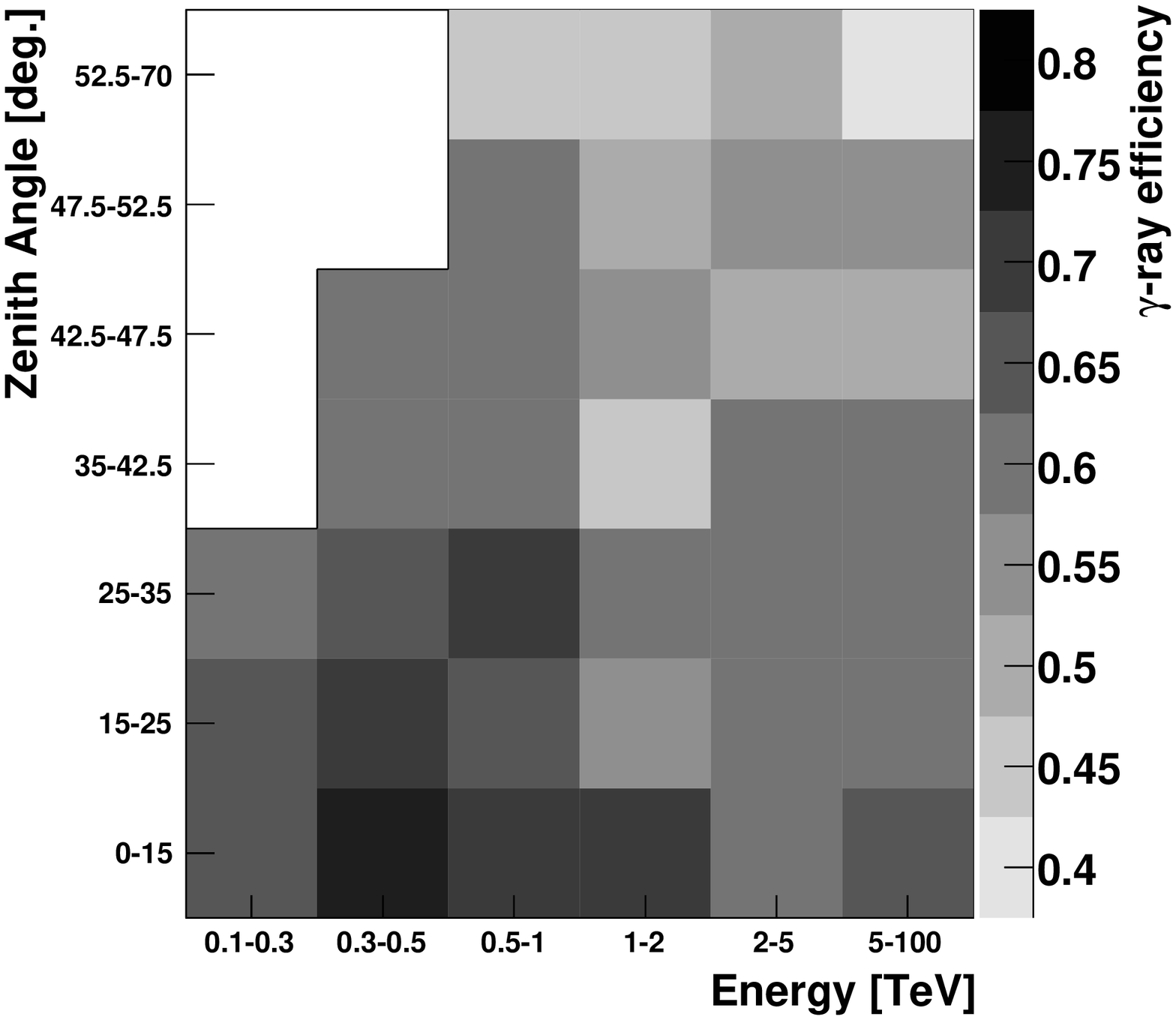}\\
  \includegraphics[width=0.45\textwidth]{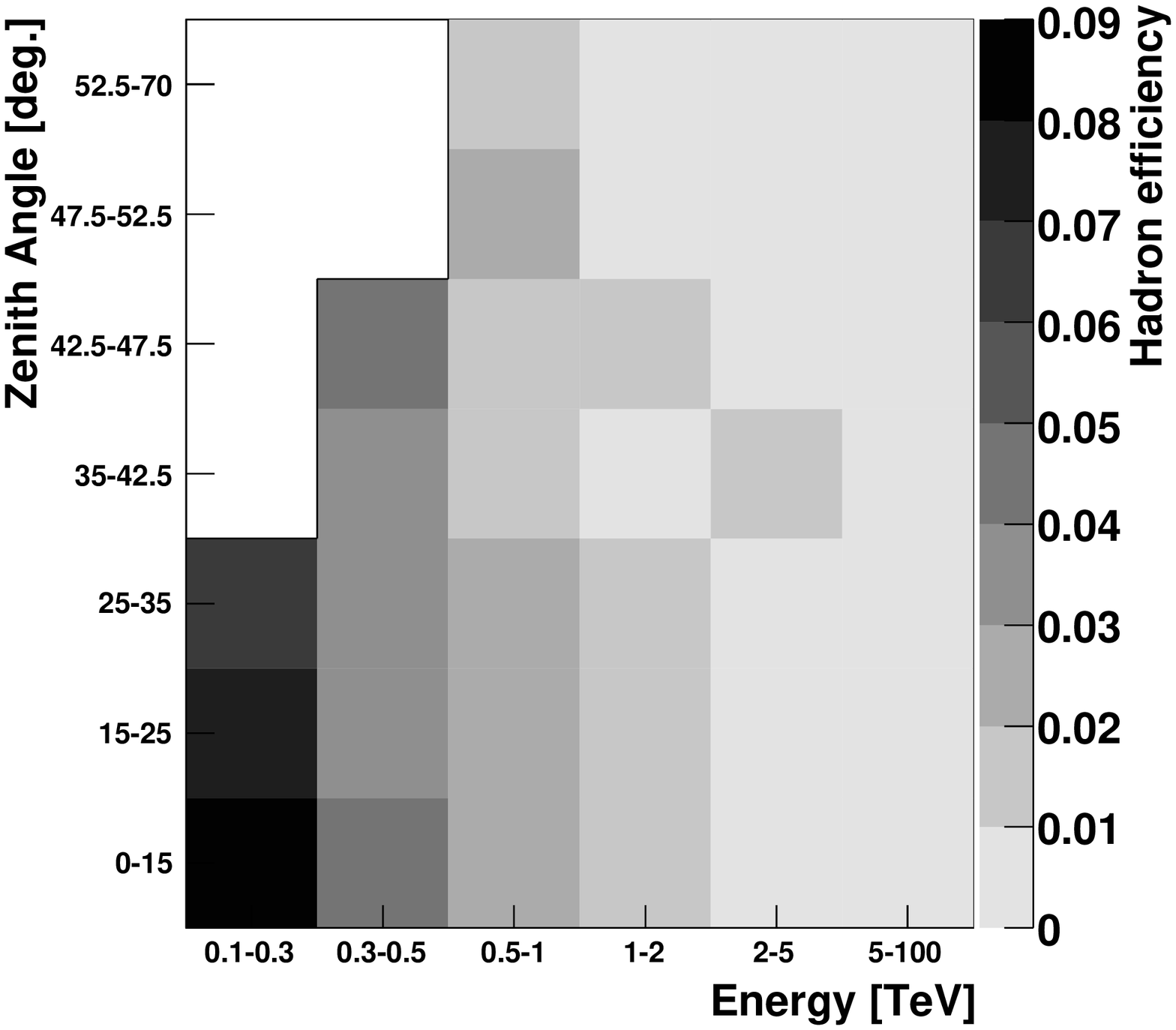}	
  \caption{{\textit Top :}~Gamma-ray efficiency distribution over the zenith angle and energy bins for the faint source set of cuts. {\textit Bottom :}~Corresponding background efficiency distribution.}
  \label{fig4}
\end{figure}

The distributions of the \gr\ efficiency and background efficiency for the chosen BDT value are shown in figure~\ref{fig4} in the case of the faint source set of cuts. The average value of the \gr\ and background efficiencies are 60\% and 2\% respectively. These efficiencies show significant dependency on the energy and zenith angle. The optimization of the analysis implies a slightly lower \gr\ efficiency at high zenith angles while the background efficiency is increased at low energies.

\section{Systematic studies}

\subsection{Comparison between Monte Carlo simulations and data}

\begin{figure}[!t]
  \centering
  \includegraphics[width=0.45\textwidth]{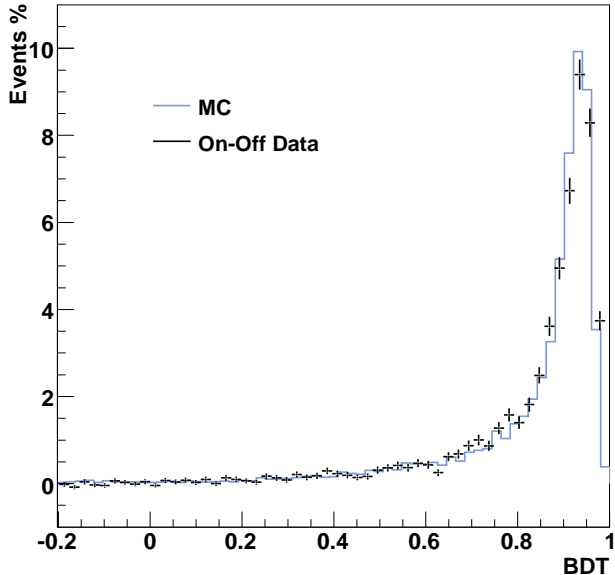}
  \caption{BDT output distributions for Monte Carlo \gr\ events and real \hess\ data with a reconstructed energy ranging between 500~GeV and 1~TeV and a zenith angle between 25\deg\ and 35\deg . The blue histogram is the Monte Carlo test sample applied to the BDT forest. The points are the On-Off event distribution from the Crab Nebula, PKS~2155-304 and HESS~J1745-290.}
  \label{fig5}
\end{figure}

The consistency between Monte Carlo simulations and real events has been checked. The MC-data consistency for the variables from the original methods has been tested in the frame of previous studies of these methods~\cite{Aharonian06a,Lemoine06,DeNaurois09}; they showed a good agreement of the simulations with \gr\ candidate \hess\ events. A good agreement has been observed for the four additional variables (primary depth and differences between reconstructed directions) in~\cite{Dubois09}.

The consistency of the BDT response between simulation and real \gr\ events has been checked for the various training. Several strong sources have been used to test this consistency: the Crab Nebula, PKS 2155-304 and HESS~J1745-290. These sources have been observed within a large range of zenith angles. This allows to test the BDT response over the seven zenith angle bins. Furthermore, the deep exposure and the brightness of observed sources provide enough statistics to check the MC-data consistency within all the energy bins. The BDT responses obtained with the simulations have been compared to the On source event distributions after subtraction of the Off source event distributions. The On-Off distribution corresponds to the \gr\ candidate events. In all the energy and zenith angle bins, a good agreement between the simulated \gr\ and the On-Off distribution has been observed. Figure~\ref{fig5} shows the simulation and On-Off data BDT response distribution for one bin. It shows the reliability and the robustness of this discriminating method.

\subsection{Comparison with current analysis and published results}

\begin{figure}[!t]
  \centering
  \includegraphics[width=0.45\textwidth]{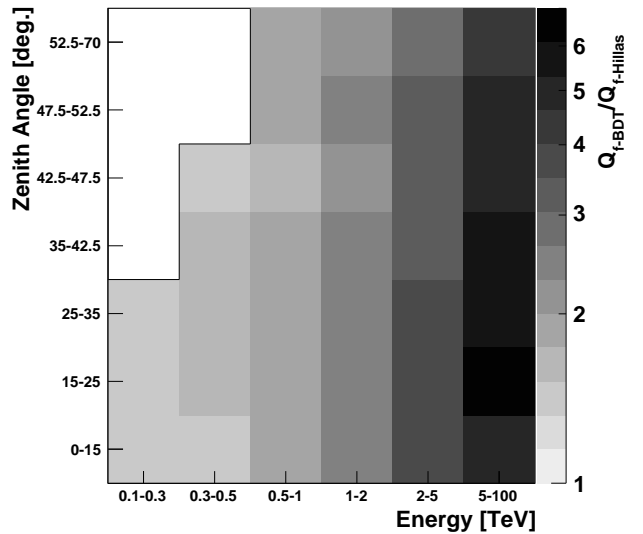}
  \caption{The figure shows the ratio between the quality factor $\mathrm{Q}_{\mathrm{f}}$ of the present work over the quality factor from the Hillas analysis soft cuts, within the zenith angle and energy bins.}
  \label{fig6}
\end{figure}

Figure~\ref{fig6} features the ratio between the quality factor of the BDT analysis and the quality factor from the Hillas analysis ($\mathrm{Q}_{\mathrm{f}}=\epsilon_{\gamma}/\sqrt{\epsilon_{\mathrm{h}}}$ where $\epsilon_{\gamma}$ and  $\epsilon_{\mathrm{h}}$ are the \gr\ and hadron efficiencies respectively). The improvement in terms of discrimination is illustrated. This ratio ranges from 1.4 to 6.6 and shows that the BDT method greatly improves the rejection in all the energies and zenith angles. The figure shows also that the improvement is a factor of the energy. As the original analyses are not energy dependant, they are mainly optimized for the lower energies, where statistics are the most important. A energy binned analysis, such as the present BDT analysis, is thus very useful to increase the discrimination power at higher energies. The effect is illustrated in the figure: at lower energy, the BDT gives a better quality factor but the major increase is located at higher energy where it reaches for some bin a value around 6.6 times the Hillas quality factor. A fraction of the improvement at high energy comes from the presence of the model and model3D methods, which are more efficient at high energy. However when compared to the model analysis, the quality factor ratio reaches values still larger than 5 at high energy. Moreover, it should be noted that due to the falling power law nature of cosmic rays, the background statistic is limited at higher energy. The major improvement of the analysis is observed above 500 GeV where the ratio is higher than 1.5. The method allows also to increase the discrimination power of the analysis at high zenith angle compared to the standard analysis.

The performance has been checked with several VHE \gr\ sources detected by \hess\ which represent a wide range of sources in term of extension, background condition (galactic or extra-galactic) and spectrum. Table~\ref{tab1} shows the results of the BDT analysis for these sources. These analyses have been made using the published data-sets. The residual background estimation has been performed using the reflected-region technique (for more details see \cite{Berge07}). The better performance of the BDT analysis compared to the original methods is illustrated on these results. The significance of the excess is greatly increased for all the sources. The signal over background ratio is, as expected, clearly increased. It allows a lower level of background contamination for subsequent spectral studies and thus sensitivity.

\begin{table*}
	\begin{center}
  	\begin{tabular}{ccccccc}
  	\hline
	Source & Method & On$^1$ & Off$^2$ & N$_{\gamma}^3$  & N$_{\sigma}^4$ & S/B$^5$  \\
 	\hline
 	\hline
    G0.9+0.1 & Hillas Hard Cuts & 1520 & 1103 & 419 & 11.6 & 0.4 \\
    & BDT & 1122 & 626 & 496 & 17.1 & 0.8 \\
 	\hline
 	\hline
  Centaurus A & Hillas Hard Cuts & 4199 & 3869 & 330 & 5.0 & 0.1 \\
    & BDT & 1517 & 1146 & 371 & 10.0 & 0.3 \\
 	\hline
 	\hline
    1ES 0347-121 & Hillas Hard Cuts & 1167 & 840 & 327 & 10.1 & 0.4 \\
    & BDT & 874 & 500 & 374 & 14.3 & 0.7 \\
 	\hline
 	\hline
    1ES 1101-232& Hillas Soft Cuts & 4276 & 3623 & 649 & 10.1 & 0.1 \\
    & BDT & 1399 & 813 & 586 & 17.9 & 0.7 \\
 	\hline
 	\hline
    H2356-309 & model3D &  1706 & 1261 & 453 & 11.6 & 0.4 \\
    & BDT & 1631 & 932 & 699 & 19.6 & 0.8 \\
 	\hline
 	\hline
    	Crab Nebula & Hillas Published & 4759 & 483 & 4283 & 94.2 & 8.8 \\
    	&Model & 10079 & 2634 & 7293 & 99 & 2.7 \\
    	&Model3D & 7460 & 1573 & 5958 & 99 & 3.8 \\
    	&BDT & 6292 & 244 & 6048 & 147.1 & 24.8 \\
 	\hline

   	\end{tabular}
  	\caption{Results obtained with the BDT analysis for various VHE \gr\ sources compared to the standard Hillas analysis or the published analysis. A comparison of the BDT analysis with the three reconstruction methods is given for the Crab. Column description: $^1$ On events $^2$ Normalised Off events $^3$ \gr\ candidates $^4$ Excess significance $^5$ Signal over background ratio.}
  	\label{tab1}
	\end{center}
\end{table*}

\subsection{Spectral analysis}

\begin{figure}[!t]
  \centering
  \includegraphics[width=0.5\textwidth]{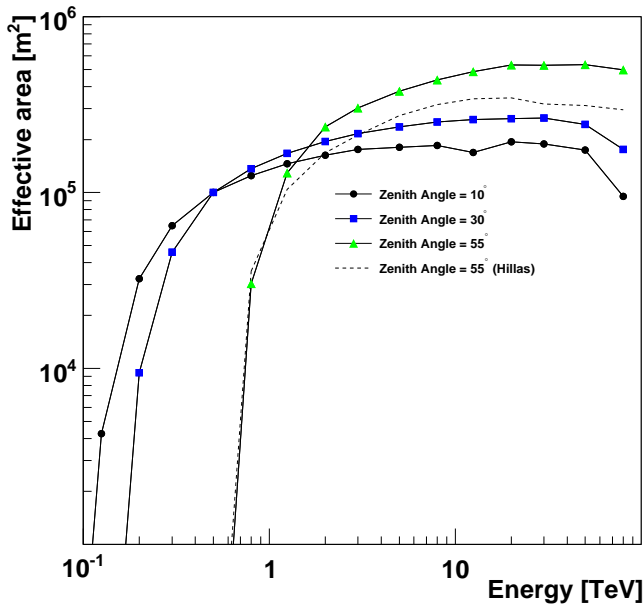}
  \caption{Effective area for \gr\ collection after application of the strong source set of cuts of the BDT method. The effective area is computed for an azimuthal angle of 180\deg\ and an off-axis angle of 0.5\deg . The black circles, blue squares and green triangles are for a zenith angle of 10\deg\, 30\deg\ and 55\deg\ respectively. The dashed line is the effective area for \gr\ collection with the Hillas method for a zenith angle of 55\deg.}
  \label{fig7}
\end{figure}

Figure~\ref{fig7} shows the energy dependency of the photon effective area for the \hess\ array as a function of the Monte Carlo simulated photon energy. The three curves are for a zenith angle of 10\deg , 30\deg\ and 55\deg . A major issue can come from the energy band optimization of the analysis. The band cut optimization can lead to a band effect within the effective area and can generate systematic fake structures within the spectrum. There is no such kind of effects visible on figure~\ref{fig7} neither on the other zenith angle bins.

The consistency of the analysis and the associated spectral analysis have also been verified on the VHE \gr\ source list from the previous section. A pure power law has been fitted to the data. Table~\ref{tab2} summarizes the spectral results obtained with the BDT analysis. They are compared to the published values. On all these reference sources, the BDT method gives consistent results with the published results. The BDT analysis allows to extend the energy range of the fit. Due to the increased rejection power from the variable combination and the fact that the optimisation of the analysis has been achieved for several bins in energy, an improvement of the analysis over the full energy range is observed. While a slight decrease of the energy threshold is observed, the gain is particularly important at higher energy where it has been shown on figure~\ref{fig6} that the discrimination has been greatly improved compared to the original methods. For instance, the energy range of the fit for the source H2356-309 has been particularly broaden at higher energies. The increase in effective area and discrimination results in keeping after selection three \gr\ events between 3~TeV and 12~TeV, that are considered as background events with the other methods. Additionally, a fit of the \gr\ spectrum in the published energy range has been performed for all these sources and gives consistent results both with the full range BDT spectra and the published spectra.

\begin{table*}
	\begin{center}
  	\begin{tabular}{ccccccc} \hline
	Source & Method & E$_{\mathrm{min}}$ & E$_{\mathrm{max}}$ & $\Gamma$ & $\Phi_{0}$ & E$_{\mathrm{cut}}$\\
 	\hline
 	\hline
    G0.9+0.1 & Pub - Hillas & 200 GeV & 9 TeV & 2.40 \mp\ 0.11 & (5.7 \mp\ 0.7)$\times10^{-12}$ $^{\dagger}$&-\\
    & BDT & 160 GeV & 12 TeV & 2.30 \mp\  0.07 & (4.5 \mp\ 0.4)$\times10^{-12}$ $^{\dagger}$&-\\
	\hline
 	\hline
    Centaurus A & Pub - Hillas  & 250~GeV & 6~TeV & 2.73 \mp\  0.45 & (2.45 \mp\  0.52)$\times10^{-13}$ &-\\
    & BDT & 200 GeV & 12 TeV & 2.71 \mp\ 0.14 & (2.32 \mp\  0.27)$\times10^{-13}$ &-\\
	\hline
	\hline
    1ES 0347-121 & Pub - Hillas & 250~GeV & 3~TeV & 3.10 \mp\ 0.23 & (4.52 \mp\ 0.85)$\times10^{-13}$ &-\\
    & BDT & 200 GeV & 4 TeV & 3.27 \mp\ 0.17 & (3.69 \mp\  0.71)$\times10^{-13}$ &-\\ 
	\hline
	\hline
    1ES 1101-232 & Pub - Hillas & 200~GeV & 4~TeV & 2.94 \mp\  0.20 & (5.63 \mp\ 0.89)$\times10^{-13}$ &-\\
    & BDT & 160 GeV & 8 TeV & 3.05 \mp\ 0.12 & (4.65 \mp\ 0.54)$\times10^{-13}$ &-\\
	\hline
 	\hline
    H2356-309 & Pub - model3D & 200~GeV & 1.1~TeV & 3.09 \mp\  0.24 & (3.00 \mp\  0.80)$\times10^{-13}$ &-\\
    & BDT &  160 GeV & 12 TeV & 3.17 \mp\  0.11 & (3.29 \mp\ 0.40)$\times10^{-13}$  &-\\
	\hline
	\hline
    Crab Nebula & Pub - Hillas & 450~GeV & 65~TeV & 2.41 \mp\  0.04 & (38.4\mp\ 0.9)$\times10^{-12}$ & 15.1 \mp\ 2.8 \\
    & model & 420~GeV & 80~TeV & 2.41 \mp\ 0.04 &  (38.2 \mp\ 0.5)$\times10^{-12}$  & 10.3 \mp\ 2.2 \\
    & model3D & 520~GeV & 75~TeV & 2.35 \mp\ 0.05 &  (35.2 \mp\ 0.8)$\times10^{-12}$  & 12.3 \mp\ 2.3 \\
    & BDT & 430 GeV & 45~TeV & 2.48 \mp\ 0.04 &  (39.0 \mp\ 0.6)$\times10^{-12}$  & 13.8 \mp\ 2.8 \\
	\hline

    \end{tabular}
  	\caption{Results of the spectral analysis performed on various VHE \gr\ sources, compared to the published values~\cite{Aharonian05a,Aharonian07a,Aharonian09,Aharonian07b,Aharonian06b}. The energy range of the fit is indicated in the first and second column, as well as the fit best parameters in the following columns. The last column is the differential flux at 1~TeV (in unit of photons~\diff).
	$^{\dagger}$ for G0.9+0.1, the last column is the integrated flux over 200~GeV (in unit of photons~\integ).}
  	\label{tab2}
	\end{center}
\end{table*}

The stability of the analysis in terms of spectral reconstruction has been tested. A variation of the three sets of cuts has been made around the nominal values.  A set of cuts optimized for a signal over background equivalent to 0.5\% and 2\% Crab Nebula has been defined for the faint source set of cuts, as well as sets at 5\% and 10\%, and 50\% and 200\%, respectively for the intermediate and strong source set of cuts. The stability of the results has been tested under these cut modifications. The tests have been performed on references sources representative of faint, intermediate and strong sources. The spectral results obtained with the modified cuts are in very good agreement with those obtained with the nominal sets of cuts. Table~\ref{tab3} summarizes the spectral results obtained for the faint source set of cuts. An additionnal test has been performed, applying the three set of cuts (faint, intermediate and strong source) on these three sources, whatever their strength. The results of this test is illustrated on table~\ref{tab3}. The \gr\ event statistics is indeed modified by the choice of cuts, but the spectral results remains unchanged. The spectral results obtained with the BDT appear very robust under cut variations, whatever the set of cuts chosen.

\begin{table}
	\begin{center}
  	\begin{tabular}{ccccc} \hline
	Source 	& Cut set & $\Gamma$ & $\Phi_{i>200\mathrm{GeV}}$ \\
		& (Crab S/B) & & \\
 	\hline
 	\hline
             & 0.5\% & 2.30 \mp\ 0.07 & 4.5 \mp\ 0.4  \\
    G0.9+0.1 & 1\% & 2.30 \mp\ 0.07 & 4.5 \mp\ 0.4 \\
             & 2\% & 2.30 \mp\ 0.07 & 4.4 \mp\ 0.4 \\
             & 10\% & 2.30 \mp\ 0.07 & 4.4 \mp\ 0.4 \\
             & 100\% & 2.32 \mp\ 0.06 & 4.7 \mp\ 0.3 \\
 	\hline
    \end{tabular}
  	\caption{Variation of spectral index from the fit under variations of the set of cuts for G0.9+0.1. The integrated flux over 200~GeV $\Phi_{i>200\mathrm{GeV}}$ is expressed in unit of $10^{-12}$ photons~\integ .}
  	\label{tab3}
	\end{center}
\end{table}

\subsection{Morphological analysis}

\begin{figure}[!t]

  \centering
   \includegraphics[width=0.4\textwidth]{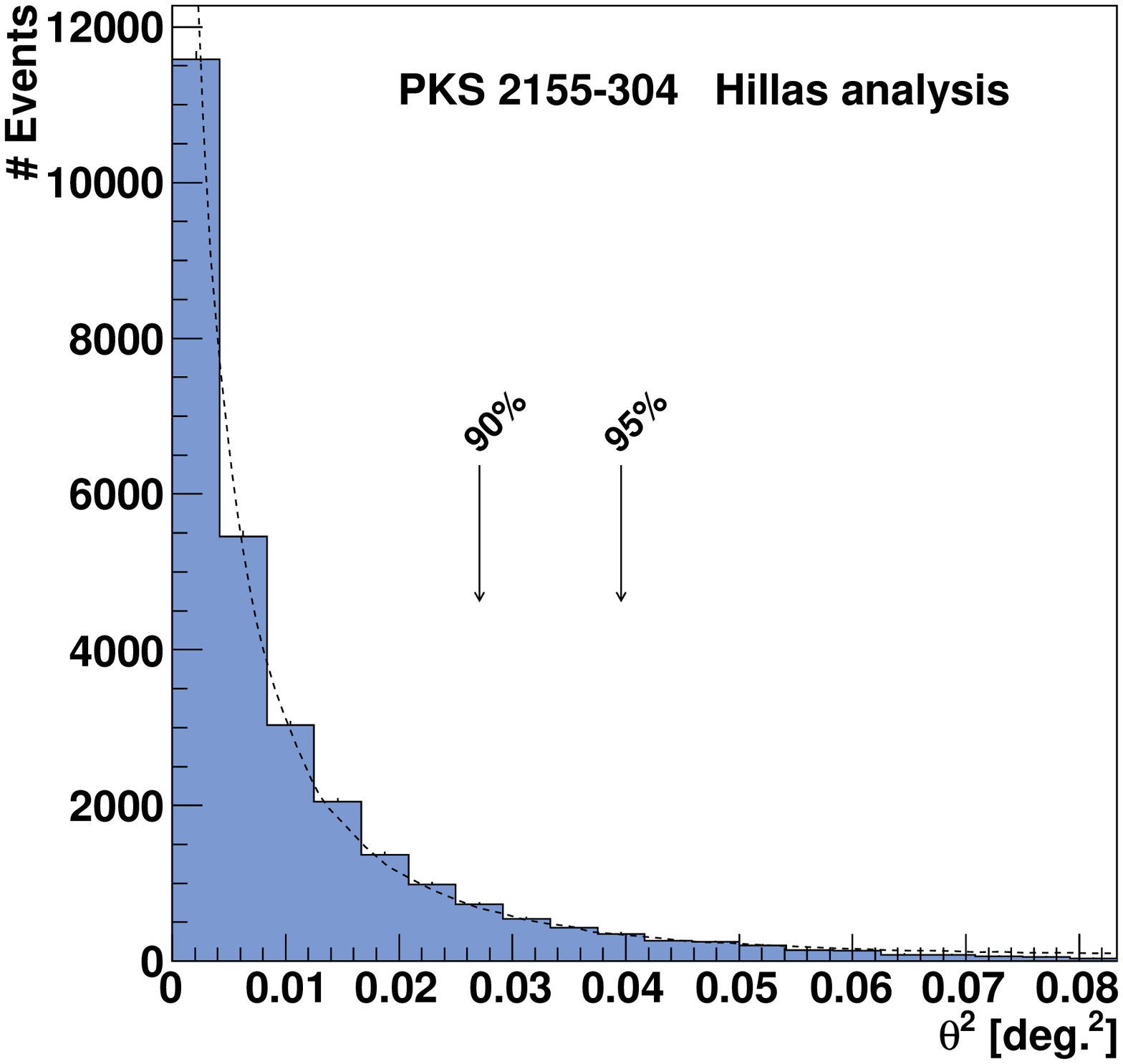}\\
   \includegraphics[width=0.4\textwidth]{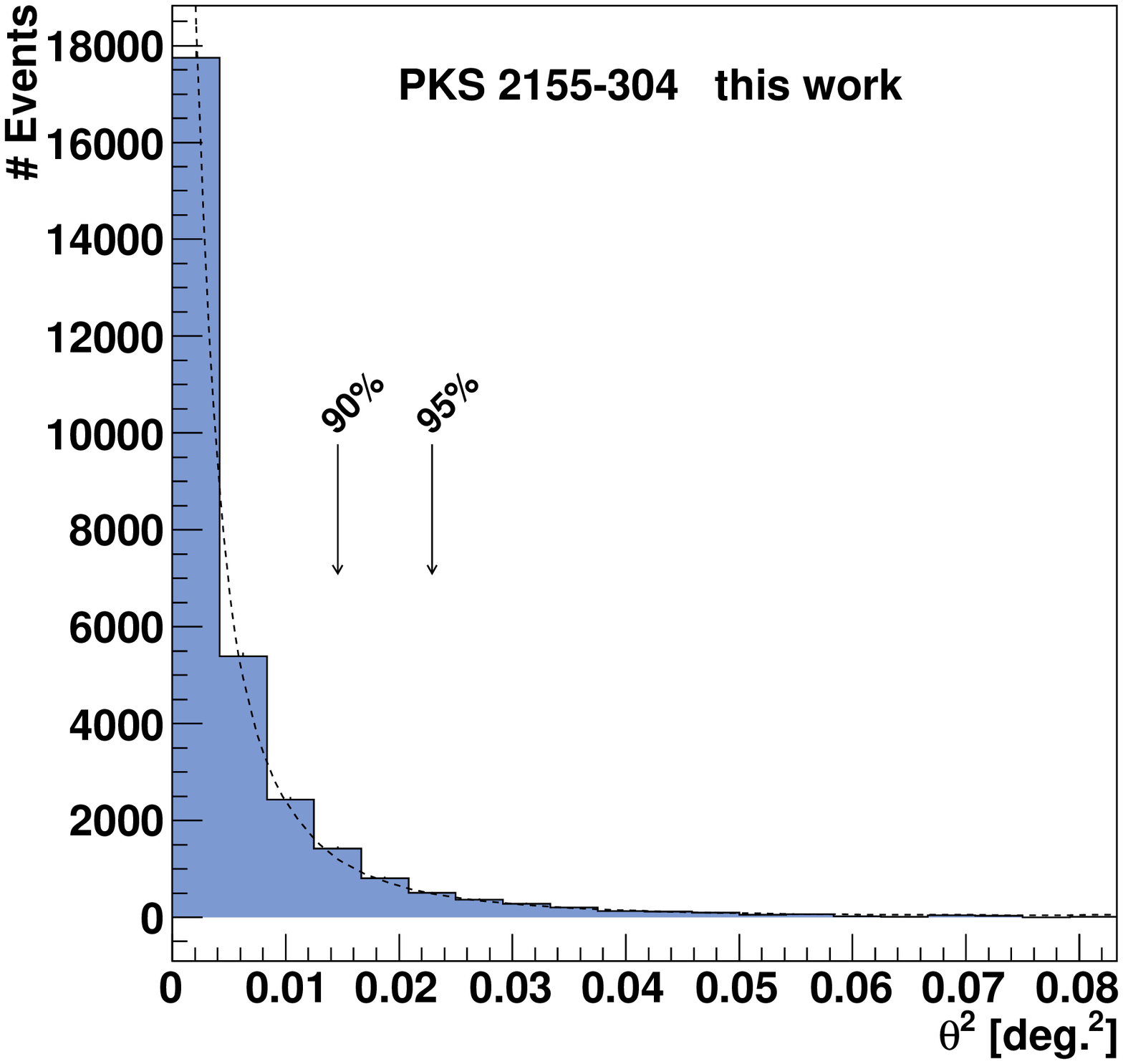}
  \caption{On-Off event distribution after selection from PKS~2155-304. The zenith angle of the shown data set ranges between 10\deg\ and 50\deg. The upper figure is obtained with the standard Hillas analysis. The lower figure is obtained with the boosted decision tree analysis described in this work. The arrows on the figures indicate the radius including 90\% and 95\% of the \gr\ signal. The dashed lines are the point spread functions obtained from Monte-Carlo simulations and corresponding to the analysis.}
  \label{fig8}
\end{figure}

The energy and direction of the selected gamma events are the combination of their corresponding estimates from each of the three standard reconstruction methods. The approach applied here, which takes into account the covariance matrices between estimates, is the same as already applied in Xeff (see~\cite{Dubois09} and reference therein for more details). It has been shown that this method gives more accurate reconstruction and improves the angular resolution of the \hess\ data analysis. An improved discrimination helps also improving the angular resolution. Figure~\ref{fig8} illustrates the benefits of the improved discrimination and the combined reconstruction to the On-Off event distribution for PKS 2155-304. This very bright point-like source is a good candidate to test the impact of the analysis method on the analysis angular resolution. The On-Off distributions are compatible with the point spread function of the instrument of the respective analysis. This distribution can be approximated by the sum of two, one-dimensional Gaussian functions. Using the fit of this sum on the distribution, the 68\% containment radius of the signal is 0.11\deg\ for the Hillas analysis and is reduced to 0.07\deg\ with the BDT method.

\section{Summary}

The discrimination between \gr\ events and hadron induced background events is a key issue for ground based Cerenkov telescopes such as \hess . A multi-variate analysis based on boosted decision trees has been studied. Three analysis methods are currently at work for the analysis of \hess\ data. The main discriminating variables from these original methods have been combined. The discrimination has been increased including the difference between the reconstructed direction of the various methods. The boosted decision trees have been trained in several bands in zenith angle and reconstructed energy in order to improve the rejection all over the energy range of the experiment and in all the observation conditions. This leads to a sizable improvement of the sensitivity. The signal over background ratio is dramatically increased compared to the original methods. The method has been tested on several reference sources which represent a wide range of sources in term of extension, nature, and observations conditions. The improvement in term of signal over background ratio and significance of the sources is illustrated. The application of this methods results also in a broader energy range for the spectral fit of faint sources compared to the previous methods. The robustness of the analysis in term of spectral reconstruction has been also demonstrated. The improved discrimination brings also a substantial gain in the angular resolution of the analysis.

\section*{Acknowledgements}

This work has been partially funded by the French National Agency for Research (ANR - Tools for Dark Matter). The authors would like to thank the H.E.S.S. Collaboration for the technical support and the fruitfull discussions. The authors thank Prof. W.~Hofmann, spokesperson of the H.E.S.S. Collaboration for allowing us to use H.E.S.S. data in this publication.


\begin{thebibliography}{} 
\bibitem{Vincent08}Vincent~P. et al. 2003, in Proc. 28th Int. Cosmic Ray Conf. (Tokyo), 2887
\bibitem{Aharonian05}(\hess\ Collaboration) Aharonian~F.A. et al., 2005, Science, 307, 1938A
\bibitem{Aharonian06c}(\hess\ Collaboration) Aharonian~F.A. et al., 2006, Astrophys. J., 636, 777A
\bibitem{Dubois09}Dubois~F., Lamanna~G. \& Jacholkowska~A., 2009, Astropart. Phys., 32, 73
\bibitem{Hillas85} Hillas~A.M., Proceedings of the 19$^{th}$ International Cosmic Ray Conference, August 11-August 23, 1985, La Jolla, USA, vol.~3, 445
\bibitem{Aharonian06a}(\hess\ Collaboration) Aharonian~F.A. et al., 2006, Astron. \& Astrophys., 457, 899-915
\bibitem{LeBohec98}Le~Bohec~S. et al., 1998, NIM A, 416, 425
\bibitem{DeNaurois09} de Naurois, M. and Rolland, L., 2009, Astropart. Phys., 32, 231 
\bibitem{Lemoine06} Lemoine-Goumard~M., Degrange~B., Tluczykont~M., 2006, Astropart. Phys., 25, 195
\bibitem{Naumann09}Naumann-God{\'o} M., Lemoine-Goumard M. \& Degrange B., 2009, Astropart. Phys., 31, 421
\bibitem{Breiman84}Breiman~L., Freidman~J.H., Olshen~R.A. \& Stone~C.J., 1984,  Classification and Regression Trees, Wadsworth ed.
\bibitem{Freund95}Freund ~Y., Schapire~R.E., 1995, Journal of Computer and System Sciences, 55
\bibitem{Aharonian08}(\hess\ Collaboration) Aharonian F.A. et al., 2008, Phys. Rev. Letters, 101, 26
\bibitem{Ohm09}Ohm~S., van Eldik~C. \& Egberts~K., 2009, Astropart. Phys., 31, 383
\bibitem{Kertzman94}Kertzman~M.P. \& Sembroski~G.H., 1994, Nuc. Inst. \& Met. in Phys. Res. A, 343, 629
\bibitem{Hoecker07}Hoecker~A. et al., 2007, arXiv:0703039
\bibitem{Berge07}Berge~D., Funk~S., \& Hinton~J.A., 2007, A\&A, 446, 1219
\bibitem{Aharonian05a}(\hess\ Collaboration) Aharonian~F.A. et al., 2005, Astron. \& Astrophys., 432L, 25A
\bibitem{Aharonian07a}(\hess\ Collaboration) Aharonian~F.A. et al., 2007, Astron. \& Astrophys., 473L, 25A
\bibitem{Aharonian09}(\hess\ Collaboration) Aharonian~F.A. et al., 2009, Astrophys. J., 695L, 40A
\bibitem{Aharonian07b}(\hess\ Collaboration) Aharonian~F.A. et al., 2007, Astron. \& Astrophys., 470, 475A
\bibitem{Aharonian06b}(\hess\ Collaboration) Aharonian~F.A. et al., 2006, Astron. \& Astrophys., 455, 461A

\end{thebibliography}
\end{document}